\documentclass[pre,aps,twocolumn,showpacs,a4paper,amsmath]{revtex4}
\usepackage{bm}
\usepackage[dvips]{graphicx}
\usepackage{amsmath,amssymb}
\usepackage{verbatim}
\begin{document}
\newcommand{\be}{\begin{equation}}
\newcommand{\en}{\end{equation}}
\newcommand{\n}{{\bf n}}
\newcommand{\m}{{\bf m}}
\newcommand{\p}{{\bf p}}
\newcommand{\br}{{\bf r}}
\newcommand{\bk}{{\bf k}}
\newcommand{\bcr}{b_{\rm cr}}
\newcommand{\lnv}{\langle \delta n^2(E)\rangle}
\newcommand{\alnv}{\langle \delta n^2\rangle}
\newcommand{\mnl}{\langle n(E)\rangle}
\newcommand{\avn}{\langle n\rangle}
\newcommand{\re}{{\rm Re}\,}
\newcommand{\erfi}{\,{\mathrm{erfi}}}

\title{Level number variance and spectral compressibility in a critical two-dimensional random matrix model}

\author{A. Ossipov$^1$, I. Rushkin$^1$, E. Cuevas$^2$}
\affiliation{$^1$School of Mathematical Sciences, University of Nottingham, Nottingham NG7 2RD, United Kingdom\\
$^2$Departamento de F\'{\i}sica, Universidad de Murcia, E-30071 Murcia, Spain}

\date{\today}

\begin{abstract}
We study level number variance in a two-dimensional random matrix model characterized by a power-law decay of the matrix elements. The amplitude of the decay is controlled by the parameter $b$.  We find analytically that at small values of $b$ the level number variance behaves linearly, with the compressibility $\chi$ between $0$ and $1$, which is typical for critical systems. For large values of $b$, we derive that $\chi=0$, as one would normally expect in the metallic phase. Using numerical simulations we determine the critical value of $b$ at which the transition between these two phases occurs.
\end{abstract}



\maketitle

\section{Introduction}

Random matrix models whose eigenstates exhibit a transition from extended to localized states provide an efficient tool for studying the Anderson metal-insulator transition \cite{Mut93, MNS94, MFD96, KM97, FOR09, PPTW11, EM08}. Their main advantage, compared to the original Anderson model \cite{And58}, is that the transition occurs not just at a single point in the parameter space, but rather on a critical line described by the variation of an additional parameter, such as the band width of the power-law banded random matrix model \cite{MFD96}. The models are accessible to perturbative treatment when this parameter is either large or small \cite{ME00, CK07, KOY11, ROF11, MG10}.

So far, most of the critical random matrix models which have been studied intensively are one-dimensional. Their Hamiltonians describe random hopping of a particle on a one-dimensional lattice in a random on-site potential.
The existence of the Anderson transition in such one-dimensional systems is related  to the long-range nature of the hopping amplitudes. In our recent work \cite{ORC11}, we studied the scaling of the moments of the eigenstates in a {\it two-dimensional} generalization of the power-law banded random matrix model \cite{PS02,C04}. In this ensemble, the matrix elements of the Hamiltonian $H_{\m\n}$ are complex independent Gaussian random variables, whose mean values are equal to zero and whose variances are determined by the distance between sites of a two-dimensional lattice:
\be\label{Ham}
\left\langle|H_{\m\n}|^2\right\rangle\equiv \frac{1}{1+(|\m-\n|/b)^4},
\en
where $\m=(m_x,m_y),\;\n=(n_x,n_y),\;1\le m_{\alpha},n_{\alpha}\le L$ are two-dimensional vectors representing two sites on a two-dimensional square lattice of size $L\times L$, and $b$ is a parameter of the model. Thus the Hamiltonian is represented by a random $L^2\times L^2$ matrix.

One natural way to convince oneself that this system is critical is to study how the moments of its eigenfunctions $\psi_n(\br)$ scale with the system size $L$. Namely, we define
\be\label{mom_def}
I_q=L^{d}\left\langle|\psi_n({\bf r})|^{2q}\right\rangle\propto L^{-d_q(q-1)},
\en
where $d$ is the dimensionality of the space and the averaging is performed over the ensemble as well as over a small energy window. Trivial exponent values $d_q=0$ and $d_q=d$ signify localized and extended states, respectively. For critical states, $0<d_q<d$ and what is more, $d_q$ generally depends on $q$, indicating that the eigenfunctions are multifractal. The above scaling of the moments of the eigenfunctions was obtained analytically and confirmed by numerical simulations in various critical random matrix ensembles \cite{EM08}. In particular, it was found that in the power-law banded random matrix model $d_q\ll 1$ when the bandwidth $b\ll 1$ (strong multifractality) and $d-d_q\ll1$ when  $b\gg 1$ (weak multifractality) \cite{MFD96,ME00}.

One of the surprising features of the model (\ref{Ham}) revealed in \cite{ORC11} was the absence of a pure power-law scaling of the moments of the eigenfunctions at $b\gg 1$. Instead of Eq.(\ref{mom_def}), the scaling was surmised to be
\be\label{ln_scaling}
I_q\propto L^{-2(q-1)}\ln^{\nu_q(q-1)} L,
\en
where the exponents $\nu_q$ play the role of the anomalous fractal dimensions $d-d_q$ and can be calculated perturbatively in a similar way \cite{ORC11}. In the same time, the standard power-law scaling (\ref{mom_def}) with $d_q\ll 1$ was found at $b\ll 1$, in full analogy with the one-dimensional version of the model.

The existence of two different scaling laws at large and small values of $b$ suggests that there should be a critical value of $b=\bcr$ separating these two regimes. One of the aims of the present paper is to show that such a critical value of $b$ does exist. While the model can be treated analytically at $b\ll 1$ or $b\gg 1$, the existence of the transition between two regimes can be investigated only with the help of numerical simulations.  These are more efficient for spectral properties rather than for the statistics of the eigenvectors. For this reason, in the present work we focus on studying the spectral compressibility.

The spectral compressibility $\chi$ is defined by the asymptotic behavior of the level number variance:
\be\label{comp-def}
\lnv=\langle n^2(E)\rangle -\langle n(E)\rangle^2\approx \chi \mnl,\quad \mnl\gg 1,
\en
where $n(E)$ is the number of eigenvalues in a spectral window of the width $E$. It is well known that $\chi$ plays the role of a critical exponent: $\chi=0$ in the metallic phase, $\chi=1$ in the localized phase and $0<\chi<1$ at criticality \cite{EM08}.

Our main results concerning the behavior of the level number variance and the compressibility in the two-dimensional model (\ref{Ham}) can be formulated as follows. The asymptotic behavior of $\lnv$ is described by Eq.(\ref{comp-def}) for all values of $b$ such that $0\le b\le \bcr$. The spectral compressibility $\chi(b)$ is a monotonically decaying function of $b$ with $\chi(0)=1$ and $\chi(\bcr)=0$. The latter equation is used as the definition of $\bcr$. At small values of $b$, $\chi(b)$ can be calculated perturbatively:
\be\label{comp_small_b}
\chi(b)=1-\frac{\pi^2 b^2}{\sqrt{2}}+O(b^4).
\en

\begin{figure}[t]
\begin{center}
\includegraphics[clip=true,width=\columnwidth]{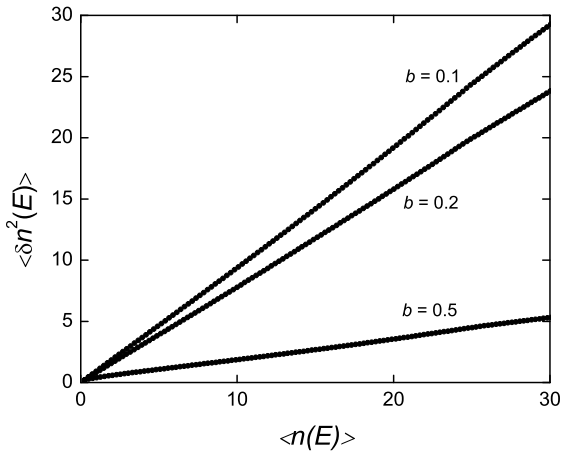}
\end{center}
\caption{Level number variance as a function of the average level number, for different values of $b<1$ and $L=64$.}
\label{Fig1}
\end{figure}

For $b\ge \bcr$, spectral compressibility is equal to zero. The level number variance in this case has the same asymptotic behavior as in the Wigner-Dyson random matrix theory:
\be
\lnv=\frac{1}{\pi^2}\left(\ln (2\pi \avn)+\gamma+1\right),\quad \avn\gg 1,
\en
where $\gamma$ is the Euler constant. Our analysis, based on numerical simulations, shows that the transition between these two regimes occurs at $\bcr=5.2 \pm 0.2$.

The remainder of this paper has the following structure. In  Section \ref{sec_small_b} we present a derivation of Eq.(\ref{comp_small_b}) at $b\ll 1$ and confirm the obtained result by numerical simulations. The opposite case of $b\gg 1$ and the transition between the two phases are considered  in Section \ref{sec_large_b}. We summarize our results in Section \ref{conclusions}. Finally, the Appendix contains an alternative derivation of Eq.(\ref{comp_small_b}), easily to extended to the orthogonal symmetry class.

\section{Spectral compressibility at $b\ll 1$}\label{sec_small_b}

A typical random matrix from the ensemble (\ref{Ham}) has the diagonal elements of order one and the off-diagonal elements of order $b$. Thus, the off-diagonal elements are parametrically smaller than the diagonal ones provided that  $b\ll 1$. This allows developing a perturbation expansion of the eigenfunction moments, spectral compressibility or any other quantity of interest in a power series of $b$. In Ref.\cite{ME00,YK03} this method was applied to the calculation of spectral compressibility in the one-dimensional power-law banded random matrix model. Below we use the same approach in the two-dimensional model (\ref{Ham}).

\begin{figure}[t]
\begin{center}
\includegraphics[clip=true,width=\columnwidth]{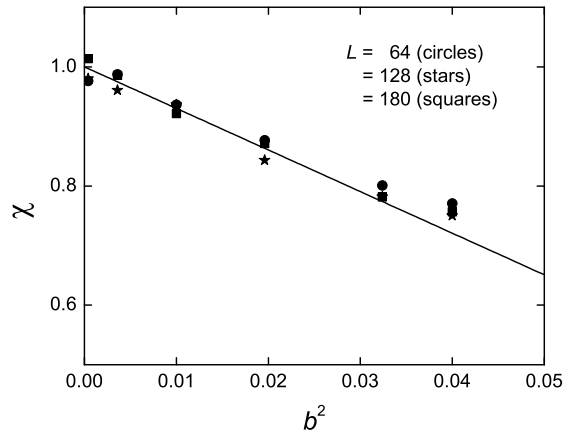}
\end{center}
\caption{Spectral compressibility as a function of $b$, obtained from the slopes of the lines in Fig.~\ref{Fig1}. The data is presented for three different system sizes. The error bars are smaller than the symbol sizes.}
\label{Fig2}
\end{figure}

The zeroth order term in the expansion corresponds to a pure diagonal random matrix, which has completely localized eigenvectors and uncorrelated eigenvalues. As a result, $\chi=1$ in this case. The first order correction to this trivial result can be found from the relation between $\chi$ and the form factor $K^{(1)}(t,N)$\cite{YK03}:
\be
\chi=1+\lim_{t\to\infty}\lim_{N\to\infty}K^{(1)}(t,N),
\en
where $N$ is the matrix size, which is equal to $L^2$ in our case, and the form factor is calculated in the lowest order of the perturbation theory. The general expression for the latter reads \cite{YK03, YO07,footnote}
\begin{eqnarray}
\label{sum_m}K^{(1)}(t,N)&=&-\frac{2\sqrt{\pi}}{t}\sum_{\m\neq 0}x(|\m|)e^{-x(|\m|)},\\
x(|\m-\n|)&=&\frac{1}{2}\langle|H_{\m\n}|^2\rangle t^2.
\end{eqnarray}
Since we are interested in the limit $N\to\infty$, the sum over $\m$ in Eq.(\ref{sum_m}) can be replaced by a two-dimensional integral, which can be transformed to the polar coordinates upon substitution of $\langle|H_{\m\n}|^2\rangle$ given in (\ref{Ham}):
\be
\lim_{N\to\infty}K^{(1)}(t,N)=-2\pi^{3/2}t\int_{1}^{\infty}dr \frac{r e^{-\frac{t^2}{1+(r/b)^4}}}{1+(r/b)^4}.
\en
This integral can be evaluated in the limit $t\to\infty$ by changing the variable from $r$ to $s=b^4t^2/2r^4$:
\be\label{correction}
\lim_{t\to\infty}\lim_{N\to\infty}K^{(1)}(t,N)=-\frac{\pi^{3/2}b^2}{\sqrt{2}}\int_{0}^{\infty}\frac{e^{-s}}{\sqrt{s}}= -\frac{\pi^{2}b^2}{\sqrt{2}},
\en
leading to the formula announced in Eq.(\ref{comp_small_b}). The same result can be obtained directly by calculating $\chi$ for a random $2\times2$ matrix, as shown in the Appendix. The corresponding expression for $\chi$ in the orthogonal symmetry class (real symmetric matrices) is also given there.

In order to test this prediction we performed numerical simulations with random matrices generated according to Eq.(\ref{Ham}). The spectra of these matrices were obtained by standard diagonalization subroutines \cite{LAPACK}, and the number of realizations for each $b$ was $20000$, $2000$ and $500$ for $L = 64$, $128$ and $180$, respectively.

The numerical findings for $b<1$ are summarized in Fig.~\ref{Fig1} and Fig.~\ref{Fig2}. The first one shows the level number variance $\lnv$ as a function of $\mnl$ for different values of $b$. The linear behavior suggested by Eq.(\ref{comp-def}) is evident for $b<1$. Extracting the slopes of the straight lines we obtain the numerical values for $\chi(b)$, which are presented in Fig.~\ref{Fig2}. They are in a good agreement  with our analytical prediction (\ref{comp_small_b}), which is valid only when the absolute value of the perturbative correction (\ref{correction}) is much smaller than one, i.e. $b^2\ll \sqrt{2}/\pi^2\approx 0.14$.

\section{Level number variance at $b\gg 1$ and the transition point}\label{sec_large_b}

To study the limit $b\gg 1$ we exploit the fact that in this case the model can be mapped onto a non-linear $\sigma$-model \cite{ORC11}. In the $\sigma$-model description the key ingredient characterizing a particular model is the propagator. In Ref.\cite{ORC11} it was found that the propagator of the two-dimensional model (\ref{Ham}) is given by
\be\label{prop}
\Pi(k)=-\frac{\pi^3}{2}\frac{1}{k^2\ln bk}
\en
in the momentum space. Its eigenvalues determine various spectral properties of the model \cite{KL95,M00}. In particular, they allow us to calculate the two-level correlation function
\be\label{R}
R(\omega)=\frac{\Delta^2}{2\pi^2}\re\sum_{\bk}\frac{1}{\frac{2 b^2}{\pi^2 \nu}k^2\ln bk-i\omega},
\en
where $\Delta$ is the mean level spacing, $\nu=1/\Delta L^2$ is the density of states and the sum runs over the discrete momenta $\bk=(2\pi m_x/L,2\pi m_y/L),\: m_{\alpha}\in \mathbb{Z}$. The two-level correlation function allows us to find the derivative of $\lnv$ \cite{KL95}
\be\label{common}
\frac{d\alnv}{d\avn}=\lim_{L\to \infty}\int_{-\avn}^{\avn}ds\: R(s), \quad s=\omega/\Delta.
\en
Substituting the expression (\ref{R}) into this formula and integrating over $s$ we obtain
\be
\frac{d\alnv}{d\avn}=\frac{\avn}{(8\pi b^2)^2}\lim_{L\to \infty}\sum_{\m}\frac{1}{m^4\ln^2\left(\frac{2\pi b m}{L}\right)+\left(\frac{\avn}{8b^2}\right)^2}.
\en
Since the sum over $\m=(m_x,m_y),\: m_{\alpha}\in \mathbb{Z}$ converges and each term with $\m\neq (0,0)$ tends to zero in the limit $L\to \infty$, we conclude that only the zero mode $\bk=0$ has a non-vanishing contribution in the thermodynamic limit.
It is well known that the contribution of the zero mode of the $\sigma$-model reproduces the results of the Wigner-Dyson random matrix theory \cite{Efetov}. Thus we must expect that in this regime $\lnv$ is given by the standard random matrix theory result \cite{Mehta}:
\be\label{rmt-res}
\lnv=\frac{1}{\pi^2}\left(\ln (2\pi \avn)+\gamma+1\right),\quad \avn\gg 1.
\en
The absence of the linear in $\avn$ term in this asymptotic law implies that $\chi=0$, as one usually finds in the metallic phase. We would like to point out that similar calculations for the one-dimensional version of the model give a non-zero result for $\chi$ \cite{MFD96, KM97}. It is the logarithmic term in the propagator (\ref{prop}), that leads to the vanishing of $\chi$. The same logarithmic term is responsible for the unusual scaling behavior of the moments of the eigenfunctions (\ref{ln_scaling}).

\begin{figure}[t]
\begin{center}
\includegraphics[clip=true,width=\columnwidth]{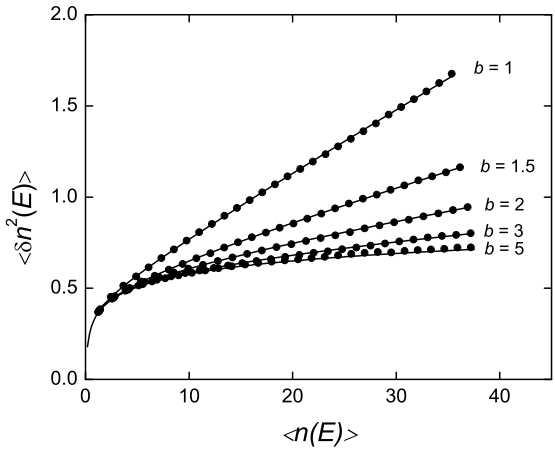}
\end{center}
\caption{Level number variance as a function of the average level number, for different values of $b>1$ and $L=64$. Dots are numerical data. The solid line at $b=5$ is Eq. (\ref{rmt-res}). The other solid lines are the results of fitting the data with Eq. (\ref{interpol}).}
\label{Fig3}
\end{figure}

Fig.~\ref{Fig3} shows the results of numerical simulations for $b\ge 1$. We observe a very good agreement between Eq. (\ref{rmt-res}) and the numeric data at $b=5$. One can also see how the behavior of $\alnv$ changes gradually from linear for small $b$ to logarithmic for large $b$. To quantify this change, as well as to determine the transition point between the two regimes, we assume that the most general functional form  of $\alnv$ is given by
\be\label{interpol}
\alnv=\chi\avn+\frac{a_1}{\pi^2}\ln(2\pi\avn)+a_2, \quad \avn\gg 1.
\en
This formula naturally interpolates between Eq. (\ref{comp-def}) and Eq. (\ref{rmt-res}). Moreover we note that Eq.(\ref{comp-def}) is never realized in its pure form: there is always a subleading logarithmic term. If the transition between the two regimes occurs at some finite value of $b=\bcr$, one should expect that $\chi$ decays as a function of $b$ and goes to zero at $\bcr$.

Using $\chi$, $a_1$ and $a_2$ as fitting parameters, we were able to reproduce the behavior of $\alnv$ at  intermediate values of $b$ as shown in Fig.~\ref{Fig3}. Moreover, the best-fit parameters for large values of $b$ are in agreement with Eq.(\ref{rmt-res}). At $b=5$, for example, we find $\chi= 0.00012 \pm 0.00016$, $a_1= 1.007 \pm 0.007$ and $a_2= 0.1573 \pm 0.003$, while the standard random matrix theory prediction (\ref{rmt-res}) corresponds to $\chi=0$, $a_1=1$ and  $a_2=0.1598$.

\begin{figure}[t]
\begin{center}
\includegraphics[clip=true,width=\columnwidth]{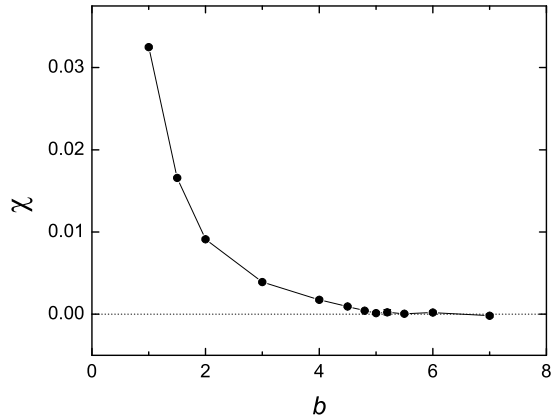}
\end{center}
\caption{Spectral compressibility as a function of $b$, extracted from the data in Fig.~\ref{Fig3} using Eq. (\ref{interpol}). The error bars are smaller than the symbol size.}
\label{Fig4}
\end{figure}

The values of $\chi$, obtained in this way for different values of $b$, are presented in Fig.~\ref{Fig4}. One can see that, indeed, $\chi$ monotonically decreases as a function of $b$ and  becomes zero within numerical accuracy for $b\ge \bcr$. A close study of $\chi$ in the interval $4.5\le b\le 5.5$ allowed us to determine the critical value $\bcr= 5.2 \pm 0.2$.

\section{Conclusions}\label{conclusions}

The random matrix model described by Eq.(\ref{Ham}) was studied numerically in Ref.\cite{PS02, C04}. The results of those works suggest that the model is critical at all values of $b$. All analytical and numerical findings known for the one-dimensional counterpart of the model support the same expectation \cite{MFD96, EM08}.

In this paper we show that the situation is actually more subtle. For $b \ll 1$ we found the expected critical behavior, characterized by a non-zero value of the spectral compressibility (\ref{comp_small_b}). Also, the scaling of the eigenstates is the standard power-law (\ref{comp-def}) with non-trivial multifractal dimensions $d_q$. For $b \gg 1$, on the other hand, we obtained $\chi=0$. This is normally a signature of the standard metallic phase corresponding to completely extended states with $d_q=d$. However, this is not the case in our model, where the moments of the eigenstates contain an additional anomalous part which scales as a power of the logarithm of system size (\ref{ln_scaling}). Therefore, the phase at $b\gg 1$ is not entirely metallic. We may say that it has some traces of the critical behavior. We found that the transition between the two phases -- ``critical" and ``metallic-critical" -- occurs at $\bcr= 5.2\pm 0.2$.

We believe that two important features of the model are responsible for the emergence of the two phases: the dimensionality $d=2$ and the long-range nature of the Hamiltonian (\ref{Ham}). Recently, a similar behavior was predicted for another two-dimensional long-range random Hamiltonian \cite{AAE11}.

Finally, we would like to point out that $\chi$ given in Eq.(\ref{comp_small_b}) and $d_1=\sqrt{2}\pi^2b^2$, as calculated in Ref.\cite{ORC11}, satisfy the relation
\be
\chi+\frac{d_1}{d}=1,
\en
which was suggested recently in Ref.\cite{BG11} and verified for various one-dimensional random matrix models.

\section{Acknowledgments}

We are grateful to Yan Fyodorov, Vladimir Kravtsov and Alexander Mirlin for useful discussions. IR and AO acknowledge support from the Engineering and Physical Sciences Research Council [grant number EP/G055769/1]. E.C. thanks the FEDER and the Spanish DGI for financial support through Project No. FIS2010-16430.

\section{Appendix. Calculation of spectral compressibility at $b\ll1$.}\label{app}

In the limit of an almost diagonal Hamiltonian $H_{ij}$ the problem can be reduced to a $2\times 2$ Hamiltonian \cite{EM08}:
\be H_2=\left(
        \begin{array}{cc}
          \epsilon_1 & b h \\
          b h^* & \epsilon_2 \\
        \end{array}
      \right),
\en
whose eigenvalues $\lambda_{\pm}$ are the solutions of the characteristic equation
\be\label{characteristic} (\epsilon_1-\lambda)(\epsilon_2-\lambda)-b^2|h|^2=0.\en Both $\epsilon_{1,2}$ are Gaussian with variance $\sigma^2$.
The density-density correlator of this $2\times 2$ matrix correlator near $E=0$ is
$R(\omega)=\frac{1}{4}\left\langle \delta(\omega-\lambda_+)\delta(\lambda_-)\right\rangle.$ We are interested in the true density correlator of the original large random matrix. Although the correlators in these two cases are related, they are not the same: it turns out that we need to double the $2\times2$ $R(\omega)$. Indeed, in a very large matrix diagonal matrix (the limit $b=0$) $\langle\rho(\omega)\rho(0)\rangle=\langle
\rho(\omega)\rangle\langle\rho(0)\rangle$, whereas in the $2\times 2$ matrix the r.h.s. contains a coefficient $1/2$. Doubling the correlator, we have
\be R(\omega)=\frac{1}{8\pi \sigma^2}\left\langle\int e^{-\frac{\epsilon_1^2+\epsilon_2^2}{2\sigma^2}}\delta(\omega -\lambda_+)\delta(\lambda_-)d\epsilon_1 d\epsilon_2 \right\rangle_h,\en
where $\langle\rangle_h$ is the average over the off-diagonal elements. Using the Eq.. (\ref{characteristic}) we can change the integration variable $\epsilon_1\to \lambda_-$. One integral is then removed by $\delta(\lambda_-).$ The other $\delta$-function ensures $\epsilon_2+\frac{b^2|h|^2}{\epsilon_2}=\omega$, which can be used in the exponential. Afterwards it becomes clear that the exponential can be neglected, as long as we are working in an energy window $\omega\ll\sigma$. In this limit,
\be R(\omega)=\frac{1}{4\pi \sigma^2}\left\langle \frac{|\omega|}{ \sqrt{\omega^2-4b^2|h|^2}}\mathbf{1}_{|\omega|>2b|h|}
\right\rangle_h.\en The average density of states at $E=0$ is given by a simpler and similar calculation:
$\langle\rho(0)\rangle=\frac{1}{\sqrt{2\pi \sigma^2}}.$
Knowing the density correlator in the $2\times 2$ case, we can find the leading order of the density correlator in the original large system by replacing $bh$ with the off-diagonal elements $H_{ij}$ and introducing the summation over $i,j$. Then we can substitute the connected correlator into Eq. (\ref{common}). As a result,
\be\label{the amazing formula} \chi=1-\lim_{\langle n\rangle\to\infty}\Bigl(2\langle n\rangle-\nonumber\en
\be\frac{2\langle n\rangle}{N^2}\sum_{i\neq j}\Bigl\langle\sqrt{1- \frac{2N^2}{\pi \langle n\rangle^2}\frac{1}{\sigma^2}|H_{ij}|^2} \mathbf{1}_{|H_{ij}|^2<\pi\sigma^2 \frac{\langle n\rangle^2}{2N^2}}\Bigr\rangle\Bigr).\en
This result is general -- valid for any dimensionality and for any symmetry class. Note that each term of the sum is an independent random variable and so can be averaged separately. Performing the averaging in the unitary case, we obtain $\chi=1-\lim_{\langle n\rangle\to\infty}I$ where
\be I^{(U)}=\sum_{\bf{r},\bf{r}'}\frac{\sqrt2a_{{\bf r}, {\bf r}'}}{N} \exp\!\left(-\frac{\pi \langle n\rangle^2}{2a_{{\bf r}, {\bf r}'}^2N^2}\right)\!\erfi\!\left(\sqrt\frac{\pi}{2}\frac{\langle n\rangle}{a_{{\bf r}, {\bf r}'}N}\right).\en
Here we used the standard notation for the variance $\langle|H_{{\bf r}, {\bf r}'}|^2\rangle=a^2_{{\bf r}, {\bf r}'}$.
In our 2D system it is sufficient to set $a_{{\bf r}, {\bf r}'}=\frac{b^2}{|{\bf r}- {\bf r}'|^2}$. To simplify the calculation, we can imagine that our system occupies a very large disk of radius $R$ (so that $N=\pi R^2$), replace summation with integration over ${\bf r}$ and ${\bf r}'$ and subsequently change the integration variables to ${\bf r}- {\bf r}'$ and ${\bf r}+{\bf r}'$. The integrals can be computed in polar coordinates. The first integration depends weakly on the large upper limit and so we can set it to $R$. Afterwards the integrand does not depend on ${\bf r}+{\bf r}'$, making the second integration trivial. The result is
\be I^{(U)}\approx 2\langle n\rangle\,_pF_q\left(\frac{1}{2},1;\frac{3}{2}, \frac{3}{2};-\frac{\langle n\rangle^2}{2\pi b^4}\right).\en
Taking the large $\langle n\rangle$ limit,
\be\chi^{(U)}=1-\frac{\pi^2 b^2}{\sqrt2}+O(b^4).\en
The case of the orthogonal symmetry is treated in precisely the same way, starting from Eq. (\ref{the amazing formula}). There we obtain
\be\chi^{(O)}=1-2\pi b^2+O(b^4).\en


\begin{thebibliography}{99}
\bibitem{footnote} The right-hand side of equation (59) in \cite{YO07} should be multiplied by $\sqrt{2}$, as in the present calculations we don't average over the energy window. Also the expression for $x(|\m|)$ contains now an additional factor $1/2$, because the variance of the diagonal elements was equal to $1/2$ in \cite{YO07}.
\end{thebibliography}
\end{document}